\documentclass[1p]{elsarticle}

\usepackage{lineno,hyperref}
\modulolinenumbers[5]

\usepackage{amsmath,graphicx,amsfonts,amssymb,epsfig,subfigure,mathrsfs,mathtools}
\usepackage{arydshln}
\usepackage{blkarray}

\usepackage{color}
\usepackage{multirow}
\usepackage{multicol}
\usepackage{lipsum}
\usepackage{rotating}
\usepackage{graphicx}%
\usepackage{algorithm}
\usepackage{algpseudocode}
\usepackage{cite}
\usepackage{tikz}
\usepackage{textcomp}
\usepackage{listings}
\usepackage{xcolor}
\makeatletter
\def\ps@pprintTitle{%
 \let\@oddhead\@empty
 \let\@evenhead\@empty
 \def\@oddfoot{}%
 \let\@evenfoot\@oddfoot}
\makeatother

\begin{document}

\begin{frontmatter}

\title{Asynchronous Parallel Computing Algorithm implemented in 1D Heat Equation with \textsf{CUDA}}

\author[mymainaddress]{Kooktae Lee\corref{mycorrespondingauthor}}
\ead{animodor@tamu.edu}

\author[mymainaddress]{Raktim Bhattacharya}
\ead{raktim@tamu.edu}

\address[mymainaddress]{Laboratory for Uncertainty Quantification\\Department of Aerospace Engineering, Texas A\&M University, College Station, TX 77843-3141, USA.}

\begin{abstract}
In this note, we present the stability as well as performance analysis of asynchronous parallel computing algorithm implemented in 1D heat equation with \textsf{CUDA}. The primary objective of this note lies in dissemination of asynchronous parallel computing algorithm by providing \textsf{CUDA} code for fast and easy implementation. We show that the simulations carried out on \textsf{nVIDIA} GPU device with asynchronous scheme outperforms synchronous parallel computing algorithm. In addition, we also discuss some drawbacks of asynchronous parallel computing algorithms.
\end{abstract}

\begin{keyword}
1D heat equation, asynchronous parallel computing algorithm, high performance computing, \textsf{CUDA}
\end{keyword}

\end{frontmatter}


\section{Introduction}
For decades, it has been reported that computing performance in parallel computation can deteriorate due to the synchronization penalty necessarily accompanied by parallel implementation of the given numerical scheme. Thus, there is a trend to relax this synchronization latency by adopting alternative approaches and techniques such as relaxed synchronization \citep{kim1998relaxed, renganarayana2012programming} or \textit{asynchronous parallel computing algorithm} \citep{bertsekas1989parallel, frommer2000asynchronous, bahi2005decentralized, fox2014parallel, lee2015async, lee2015convergence}. 
Although the asynchronous parallel computing algorithm has arisen to overcome the synchronization bottleneck, and hence speed up the computation, the randomness of asynchrony incurs unpredictability of the solution, which in turn leads to numerical inaccuracy of the solution or even instability in the worst case. Therefore, asynchronous algorithms have to be analyzed rigorously before it is fully implemented.

In \citep{lee2015async}, we have developed mathematical proofs for stability, rate of convergence, and error probability of asynchronous 1D heat equation via dynamical system framework (especially, the switched system framework \citep{lee2015stability, lee2015performance}). All the results in this note are based on our previous research works \citep{lee2015async}.
Thus, this note aims at testing asynchronous scheme in 1D heat equation with \textsf{CUDA} rather than developing theory and proof. In particular, we mainly focus on easy implementation of asynchronous algorithm by providing the \textsf{CUDA} code, to achieve high performance computing. In summary, the primary goal of this note lies in dissemination of the asynchronous parallel computing algorithm to enhance the computing performance of conventional parallel computation. For more details regarding the theoretical development, the readers may refer to \citep{lee2015async}. The simulations carried out on \textsf{nVIDIA GPU} device with \textsf{CUDA} present the stability result and performance analysis as well. In the last section, we also discuss some drawbacks of asynchronous parallel computing algorithm.

\section{Problem Formulation}

Consider 1D heat equation, of which partial differential equation (PDE) is given by
\begin{align}
\frac{\partial u}{\partial t} = \alpha\frac{\partial^2 u}{\partial x^2}, \quad t\geq 0, \label{eqn:1}
\end{align}
where $u$ is the time and space-varying state of the temperature, and $t$ and $x$ are continuous time and space respectively. The constant $\alpha>0$ is the thermal diffusivity of the given material. 

The PDE is solved numerically using the finite difference method by Euler explicit scheme, with a forward difference in time and a central difference in space. Adopting this finite difference method leads to
\begin{align*}
\frac{\partial u}{\partial t} &\approx \dfrac{u_{i}(k+1) - u_{i}(k)}{\Delta t},\\
\frac{\partial^2 u}{\partial x^2} &\approx \dfrac{u_{i+1}(k) - 2u_{i}(k) + u_{i-1}(k)}{\Delta x^2},
\end{align*}
where $k\in\{0,1,2,\hdots\}$ is the discrete-time index and $u_i$, $i=1,2,\hdots, N$, is the temperature value at $i^{th}$ grid space point with total $N$ numbers of the grid point. Thus \eqref{eqn:1} is approximated as 
\begin{align}
\frac{u_{i}(k+1) - u_{i}(k)}{\Delta t} &= \alpha\left(\frac{u_{i+1}(k) - 2u_{i}(k) + u_{i-1}(k)}{\Delta x^2}\right),\label{eqn:pde}
\end{align}
where the symbols $\Delta t$ and $\Delta x$ denote the sampling time and the grid resolution in space, respectively. Further, if we define a constant $ r\triangleq\alpha\frac{\Delta t}{\Delta x^2}$, then \eqref{eqn:pde} can be written as 
\begin{align}
u_{i}(k+1) = ru_{i+1}(k) + (1-2r)u_{i}(k) + ru_{i-1}(k),\label{eqn:sync}
\end{align}
where the parameter $r$ lies in between $0$ and $0.5$ for the numerical stability (see pp. 18, \citep{smith1985numerical}). 
Also, we consider the Dirichlet boundary condition (see pp. 150, \citep{pletcher2012computational}), i.e., the temperature at each end-point is invariant in time as follows:
\begin{align*}
u_1(k) = c_1,\qquad u_N(k) = c_2, \quad \forall k
\end{align*}
with some constants $c_1$ and $c_2$.

\begin{figure}[h!]
\begin{center}
\includegraphics[scale=0.5]{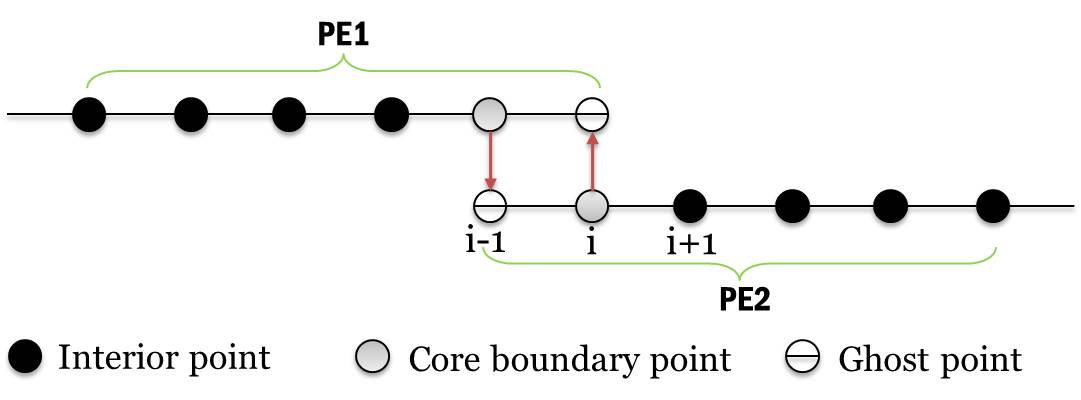}
\caption{Discretized one-dimensional domain with an asynchronous numerical algorithm. The PE denotes a group of grid points, assigned to each core.}\label{fig.0}
\end{center}
\end{figure}
Fig. \ref{fig.0} illustrates the numerical scheme over the discretized 1D spatial domain. A typical \textit{synchronous} parallel implementation of this numerical scheme  assigns several of these grid points to each processing element (PE). The updates for the temperature at the grid points assigned to each PE, occur in parallel. However, at every time step $k$, the data associated with the boundary grid points, where the communication is necessary are synchronized, and used to compute $u_i(k+1)$. This synchronization across PEs is slow, especially for massively parallel systems (estimates of idle time due to this synchronization give figures of up to 80\% of the total time taken for the simulation as idle time). 

\section{Asynchronous Parallel Computing Algorithm}
Recently, an alternative implementation -- \textit{asynchronous} algorithm -- has been proposed. In this implementation, the updates in a PE occur without waiting for the other PEs to finish and their results to be synchronized. The data update across PEs occurs sporadically and independently. This asynchrony directly affects the update equation for the boundary points, as they depend on the grid points across PEs. For these points, the update is performed with the most recent available value, typically stored in a buffer. The effect of this asynchrony then propagates to other grid points. Within a PE, we assume there is no asynchrony and data is available in a common memory. 

Thus, the asynchronous numerical scheme corresponding to \eqref{eqn:sync} is given by
\begin{align}
u_{i}(k+1) = ru_{i+1}(k^{*}_{i+1}) + (1-2r)u_{i}(k) + ru_{i-1}(k^{*}_{i-1}),\label{eqn:async}
\end{align}
where $k^{*}_{i}\in \{k,k-1,k-2,\hdots,k-q+1\}$, $i=1,2,\hdots,N$, denotes the randomness caused by communication delays between PEs. The subscript $i$ in $k_i^*$ depicts that each grid space point may have different time delays.
The parameter $q$ is the length of a buffer that every core maintains to store data transmitted from the other cores. 

In the following section, we provide the \textsf{CUDA} codes for both synchronous and asynchronous implementation of 1D heat equation.

\section{\textsf{CUDA} Code}
%

At first, we take a look at the synchronous code. In the parallel implementation of \eqref{eqn:sync}, only time-loop for index $k$ is necessary, since the space-loop for index $i$ can be carried out in parallel. To enforce synchronization, the parallel computation in space index $i$ is performed only once in \textsf{CUDA} kernel, and then we repeat this process thereafter in the $main$ through discrete-time iteration. After executing $kernel$, it is guaranteed that each \texttt{u[i]} value is computed and copied to the host memory. Thus, the synchronization is imposed at each instance. The time-loop is then given in the $main$, instead of $kernel$, for the synchronization issue. This would be a naive way to synchronize and alternative techniques can be also applied for synchronization such as `\textsf{\underbar{  }\underbar{ }syncthreads()}', which may further increase computing performance. 

\begin{itemize}
{\color[rgb]{0.54, 0.2, 0.14}{\item \bf Synchronous Algorithm}}
\begin{lstlisting}
__global__ void kernel(float* u){

    int i = blockIdx.x*blockDim.x + threadIdx.x;
  
	if( i > 0 && i < N-1){
	   u[i] = r*(u[i+1]-2*u[i]+a[i-1]) + u[i];
	}
}


int main(){
    float *u, *uDev;
    int size1 = N*sizeof(float);

    cudaMalloc((void**) &aDev,size1);
    u = (float*)malloc(size1);

   // initial condition
    for(int i=0;i<N;i++)
    {
	   u[i] = pow(cos( 3*PI/2*i/(N-1) ),2); // cosine func.
    }

    int dimThreads = Npts;
    int dimBlock = Npts/dimThreads;
    for(int k=0;k<kend;k++) // time-loop
    {
	   cudaMemcpy(uDev,u,size1,cudaMemcpyHostToDevice);
	   kernel<<<dimBlock, dimThreads>>>(uDev);
	   cudaMemcpy(u,uDev,size1,cudaMemcpyDeviceToHost);
    }
   free(u);
   cudaFree(uDev);
   return 0;
}
\end{lstlisting}

Next, we consider the asynchronous code. The major difference between synchronous and asynchronous codes are the placement of the time-loop. In this asynchronous code, the time-loop is imposed in the $kernel$, and hence each \texttt{u[i]} can be updated asynchronously without any  $barrier$ for synchronization. The purpose of new variables \texttt{`v'} and \texttt{`vDev'} in the asynchronous code is to keep track of \texttt{u[i]} in time, since one cannot save the history of \texttt{u[i]} while processing $kernel$.

{\color[rgb]{0.54, 0.2, 0.14}{\item \bf Asynchronous Algorithm}}
\begin{lstlisting}
__global__ void kernel(float* u, float* v){

    int i = blockIdx.x*blockDim.x + threadIdx.x;
  
    for(int k=0;k<kend;k++) // time-loop
    {
	if( i > 0 && i < N-1){
	   u[i] = r*(u[i+1]-2*u[i]+u[i-1]) + u[i];
	}
	v[N*k+i] = u[i];
    }
}


int main(){
    float *u, *uDev, *v, *vDev;
    int size1 = N*sizeof(float);
    int size2 = N*kend*sizeof(float);

    cudaMalloc((void**) &uDev,size1);
    cudaMalloc((void**) &vDev,size2);
    u = (float*)malloc(size1);
    v = (float*)malloc(size2);

   // initial condition
    for(int i=0;i<N;i++)
    {
	   u[i] = pow(cos( 3*PI/2*i/(N-1) ),2); // cosine func.
    }

    cudaMemcpy(uDev,u,size1,cudaMemcpyHostToDevice);
    int dimThreads = Npts;
    int dimBlock = Npts/dimThreads;
    kernel<<<dimBlock,dimThreads>>>(uDev, vDev);
    cudaMemcpy(v,vDev,size2,cudaMemcpyDeviceToHost);
 
   free(u);
   free(v);
   cudaFree(uDev);
   cudaFree(vDev);
   return 0;
}
\end{lstlisting}
\end{itemize}

\section{Simulation Results}
We implemented the asynchronous parallel algorithm with $\mathtt{CUDA\:\:C\text{++}\:\:programming}$ on $\mathtt{nVIDIA\: Tesla^{{\scriptsize{TM}}}\: C2050}$ GPU, which has $448$ $\mathtt{CUDA\: cores}$. 
The simulations were performed with the following parameters:
\begin{itemize}
\item{Simulation Parameters:}
\begin{eqnarray*}
&\Delta x =& 0.1, \Delta t = 0.01, \alpha = 0.5, r = \alpha\dfrac{\Delta t}{\Delta x^2} = 0.5\\
&I.C.:& u_i = \text{cos}^2\bigg( \dfrac{3\pi i}{2(N-1)} \bigg), \:i=1,2,\hdots,N\\
&B.C.:& u_1(k) = 1,\: u_N(k) = 0, \: \forall k
\end{eqnarray*}
\item{Number of PEs:} $N=100$.
\item{Number of grid points in PE:} $n=1$\\
\end{itemize}

In this simulation, we consider that each grid point is assigned to each PE (\textsf{CUDA} core in here) as defined $n=1$ above. Thus, each \textsf{CUDA} core updates the value of $u_i$.

\subsection{Spatio-temporal evolution of temperature}
\begin{figure}
\centering
\includegraphics[scale=0.21]{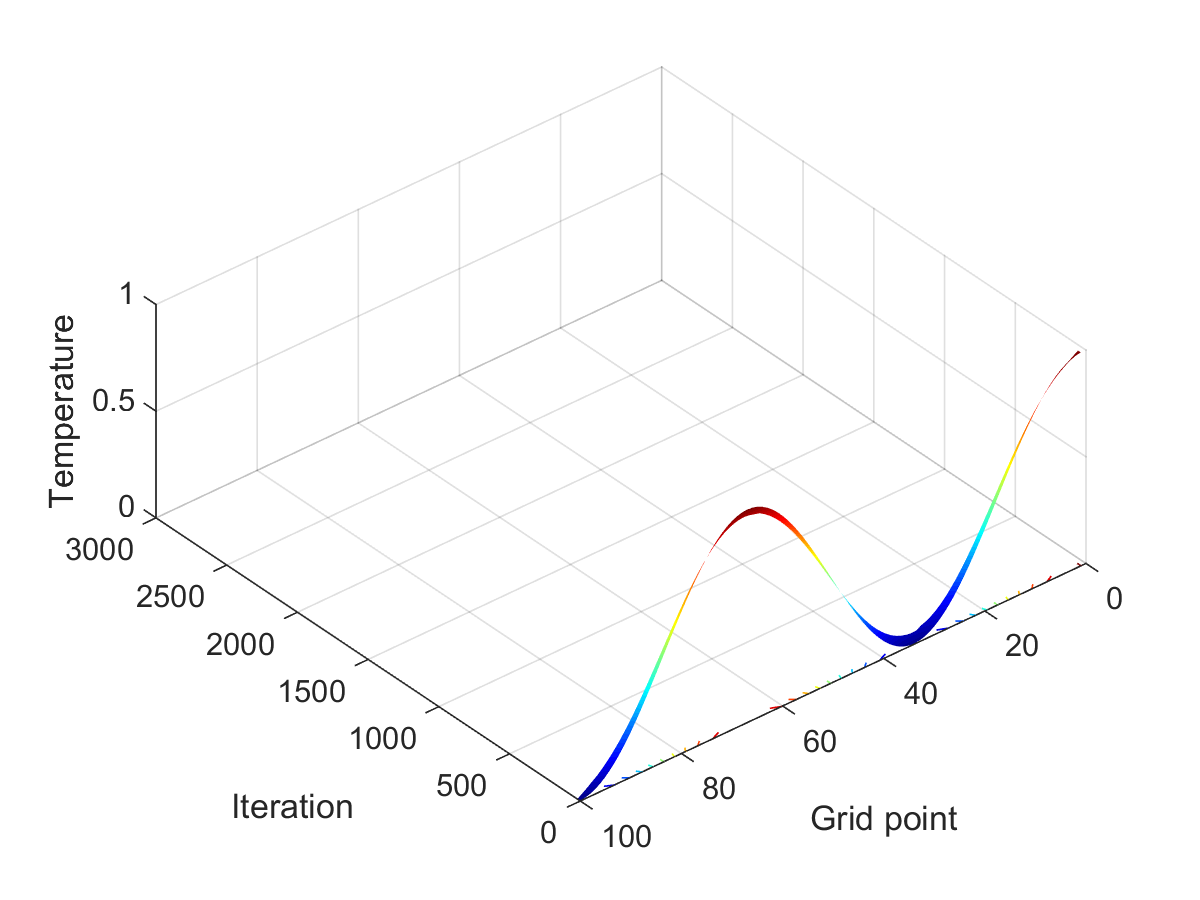}
\includegraphics[scale=0.21]{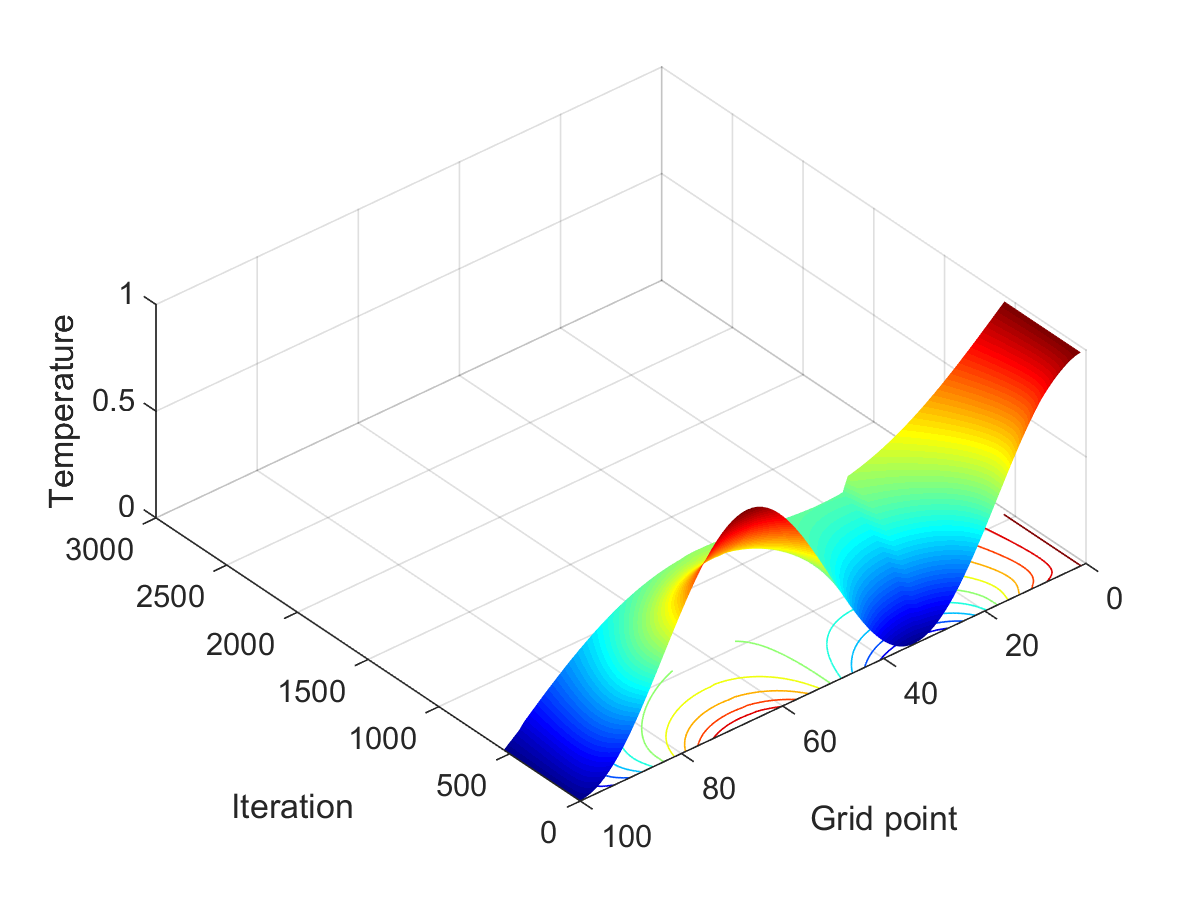}
\includegraphics[scale=0.21]{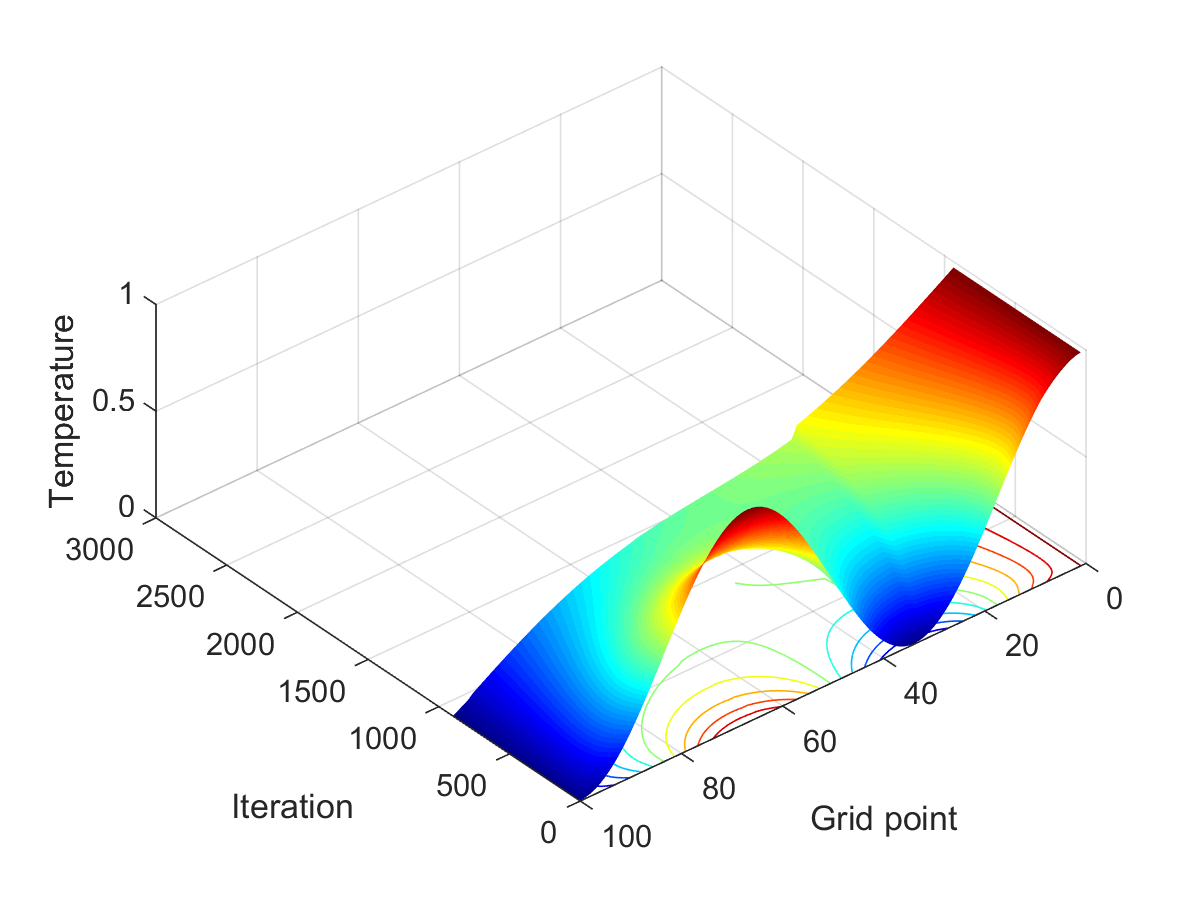}\\
(a) \qquad\qquad\qquad\qquad\qquad\quad (b) \qquad\qquad\qquad\qquad\qquad\quad (c)\\
\includegraphics[scale=0.21]{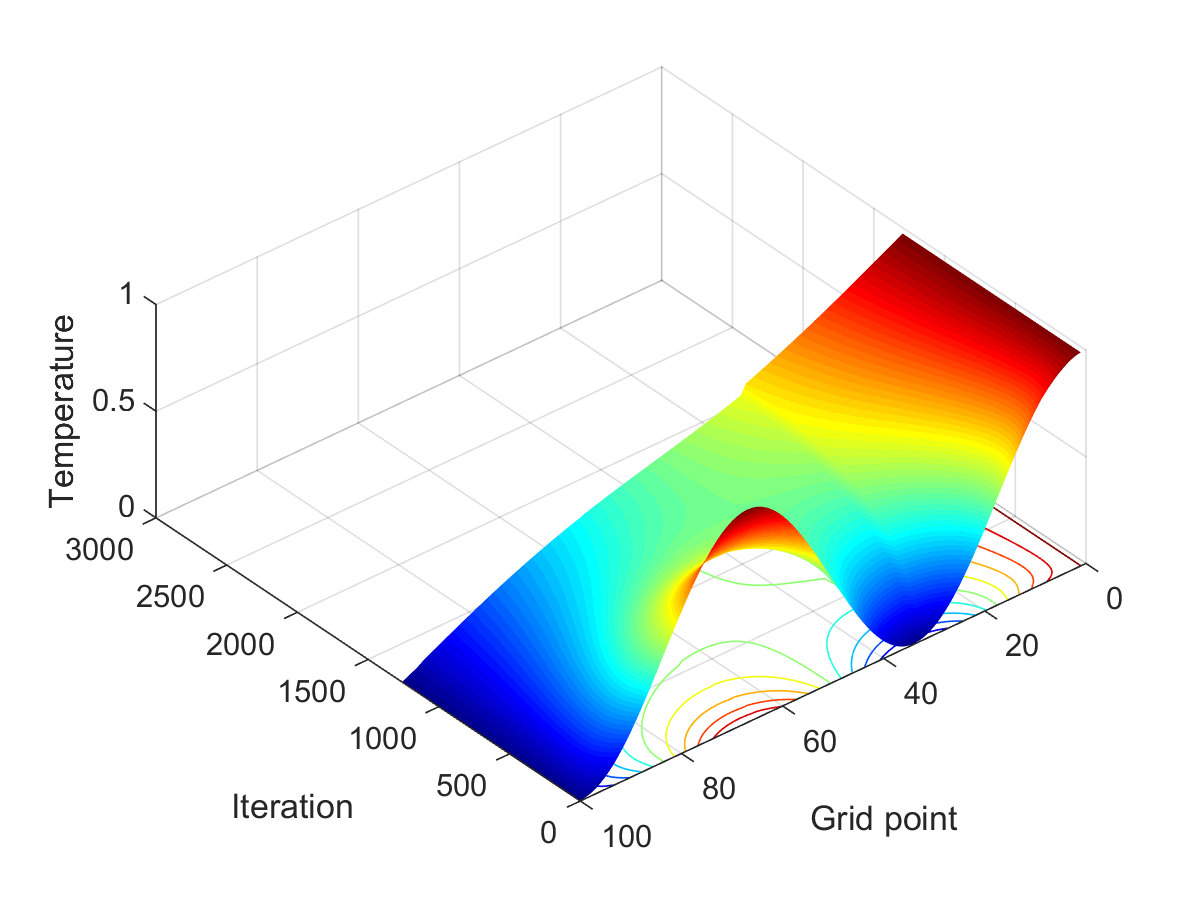}
\includegraphics[scale=0.21]{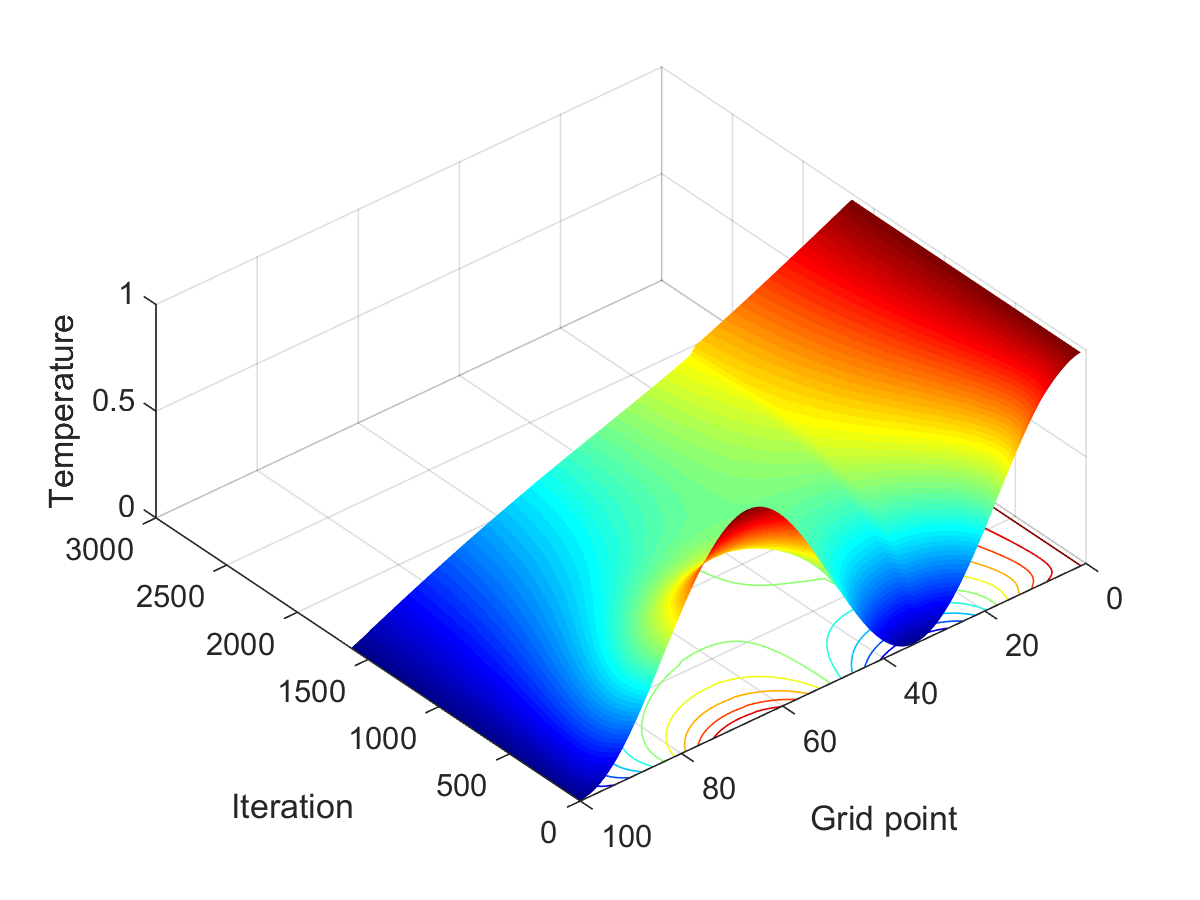}
\includegraphics[scale=0.21]{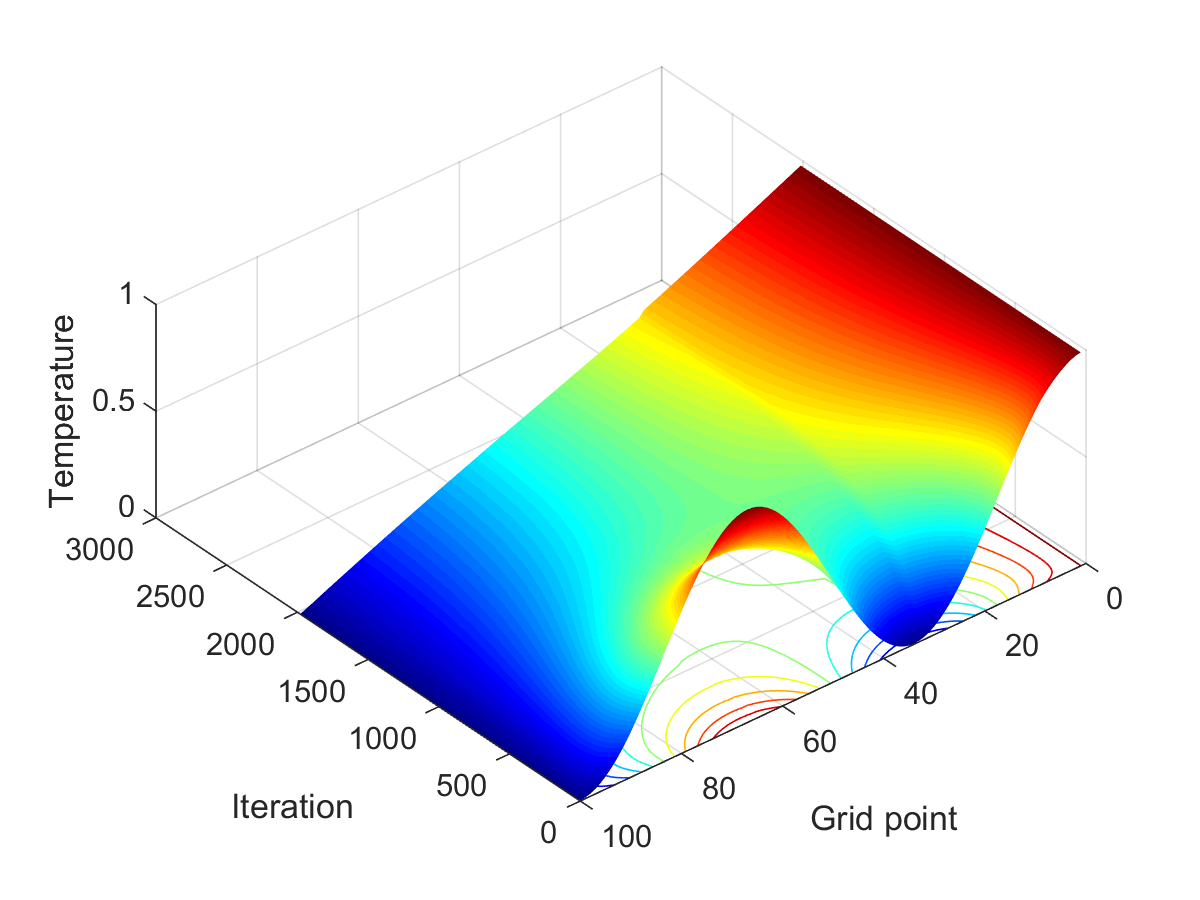}\\
(d) \qquad\qquad\qquad\qquad\qquad\quad (e) \qquad\qquad\qquad\qquad\qquad\quad (f)\\
\includegraphics[scale=0.21]{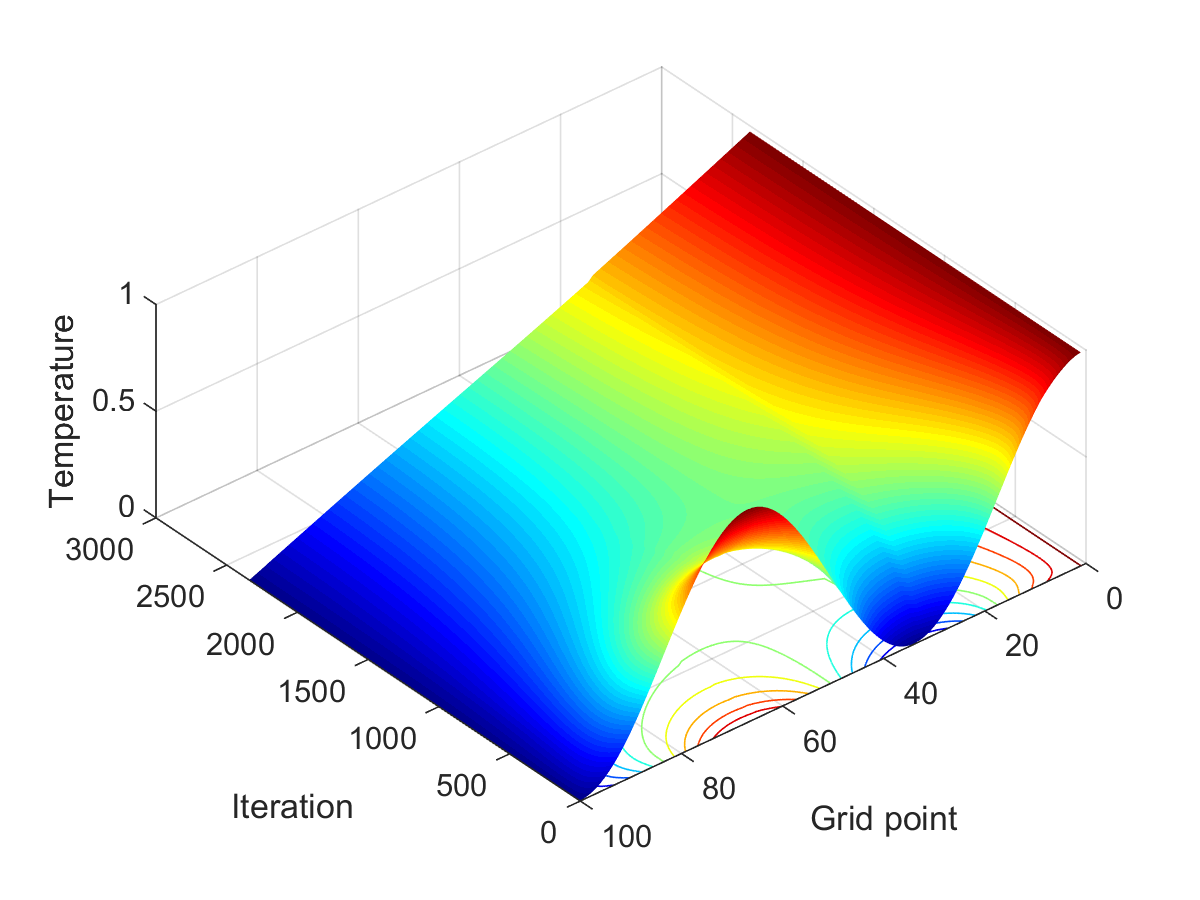}
\includegraphics[scale=0.21]{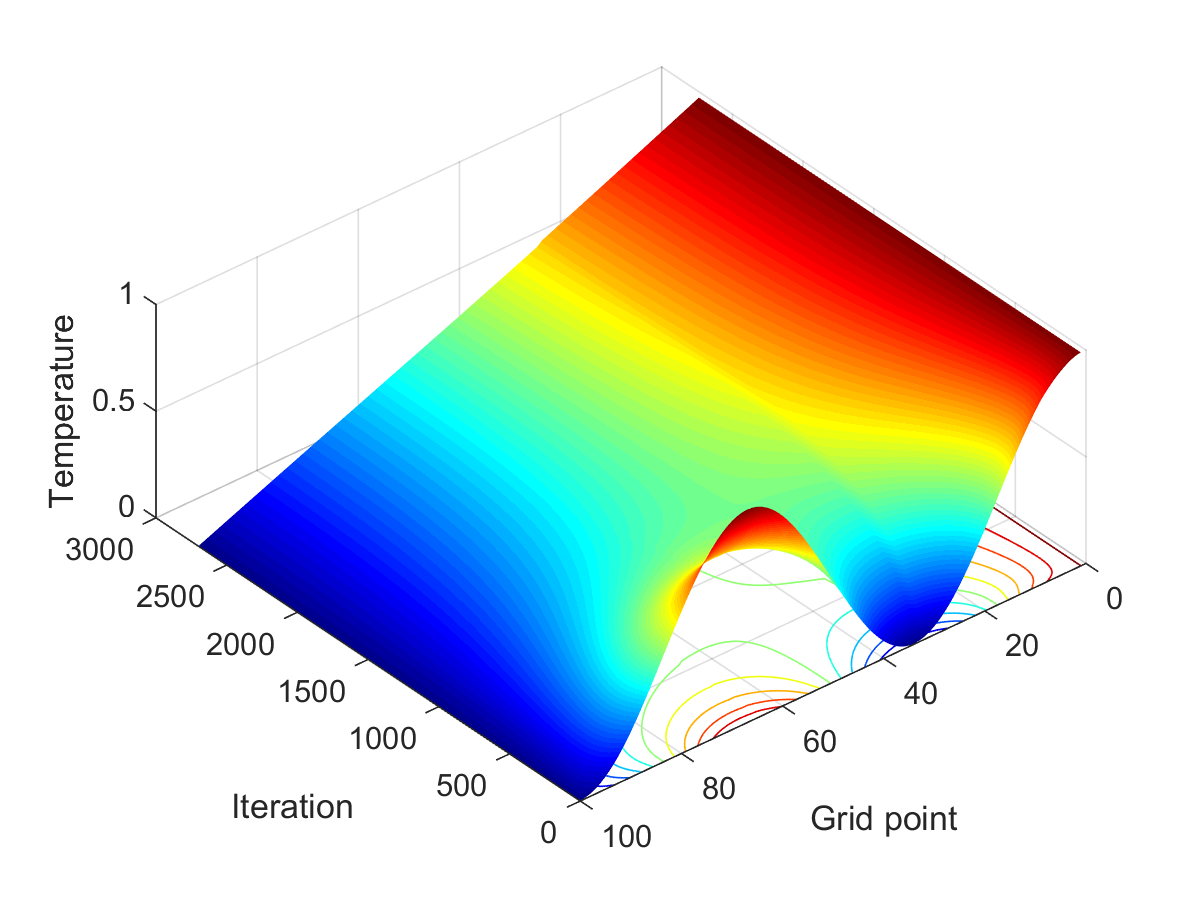}
\includegraphics[scale=0.21]{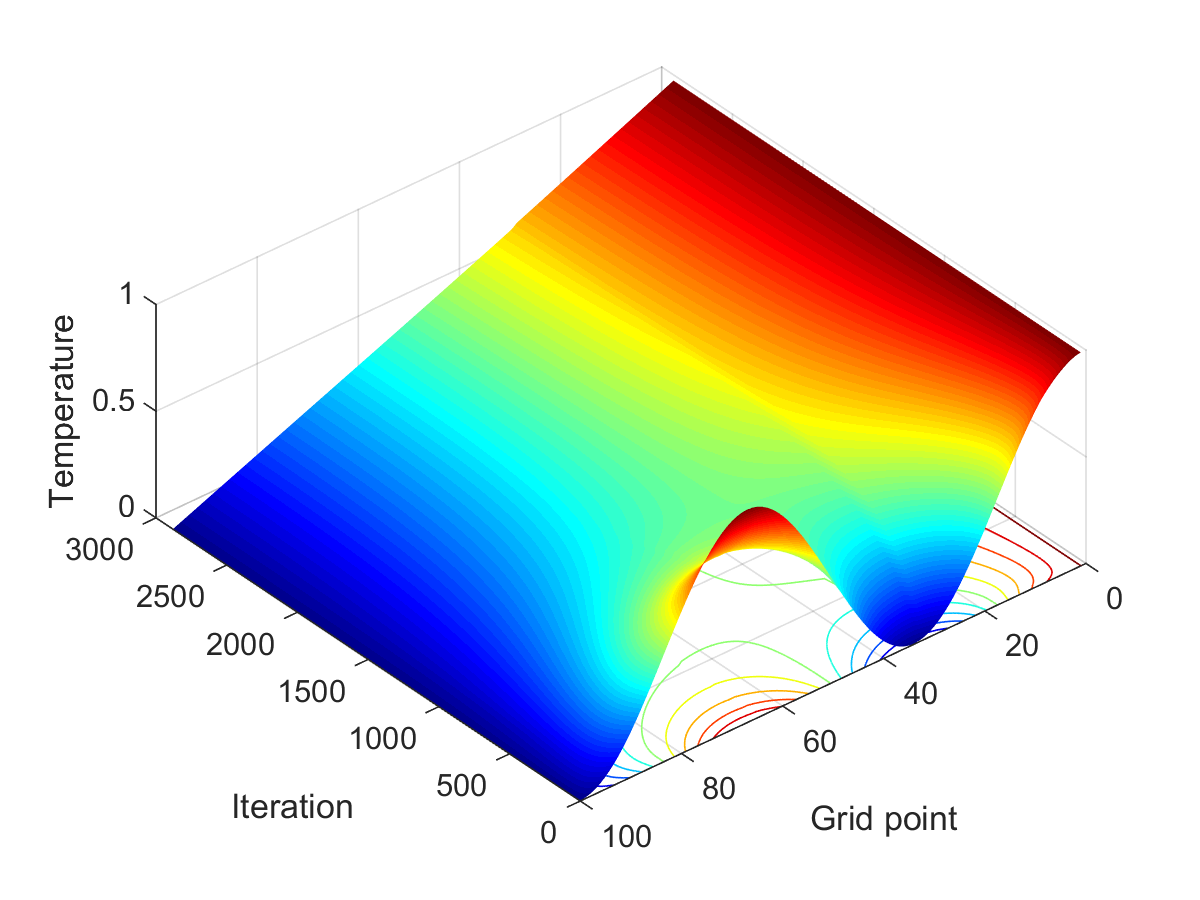}\\
(g) \qquad\qquad\qquad\qquad\qquad\quad (h) \qquad\qquad\qquad\qquad\qquad\quad (i)\\
\caption{The spatio-temporal evolution of 1D heat equation using asynchronous parallel iterative method}\label{fig.2}
\vspace{0.2in}
\end{figure}

For a given initial temperature, the spatio-temporal evolution of the state is presented in Fig. \ref{fig.2}. As time $k$ increases, the curved shape of the temperature, given as a cosine square function initially, flattens out. 

\begin{figure}
\centering
\includegraphics[scale=0.6]{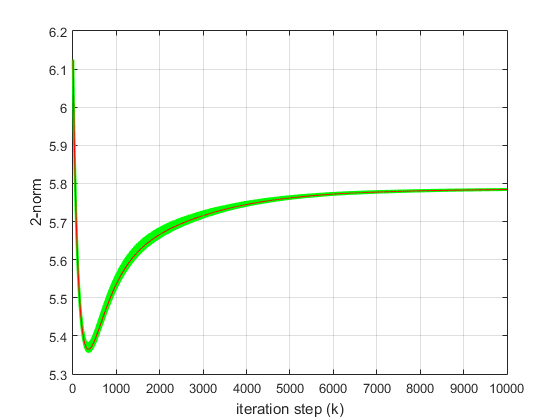}
\caption{The ensemble (green lines) of $300$ trajectories in 2-norm values with regards to the discrete temperature $u(k) = [u_1^{T}(k),\hdots, u_N^{T}(k)]^{T}$ for asynchronous scheme and their corresponding mean value (red line).}\label{fig.3}
\end{figure}

In Fig. \ref{fig.3} , the ensemble of the trajectories is shown for the asynchronous algorithm. The solid lines show the trajectories of total $300$ simulations. Due to the randomness in the asynchronous algorithm, the trajectories differ from each other. 
As we already proved in \citep{lee2015async}, the asynchronous scheme for 1D heat equation is numerically stable, which is also verified in this simulation with \textsf{CUDA}.

\subsection{Performance analysis}
\begin{figure}
\centering
\includegraphics[scale=0.4]{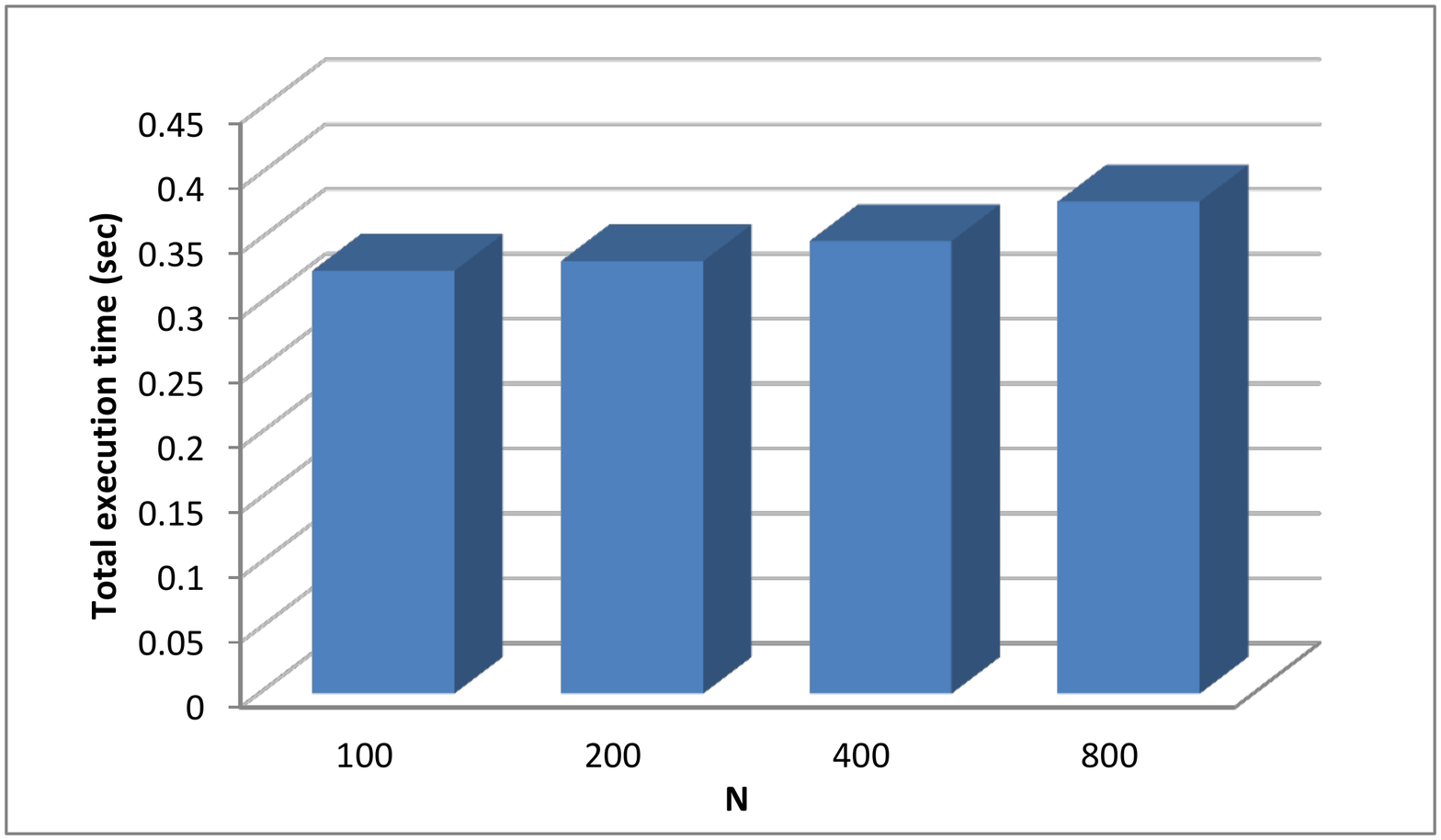}\vspace{-0.3in}
\caption{Total execution time for \textbf{\textit{synchro1nous}} parallel computing algorithm with respect to variation in the number of grid points $N$}\label{fig.4}
\end{figure}

\begin{figure}
\centering
\includegraphics[scale=0.4]{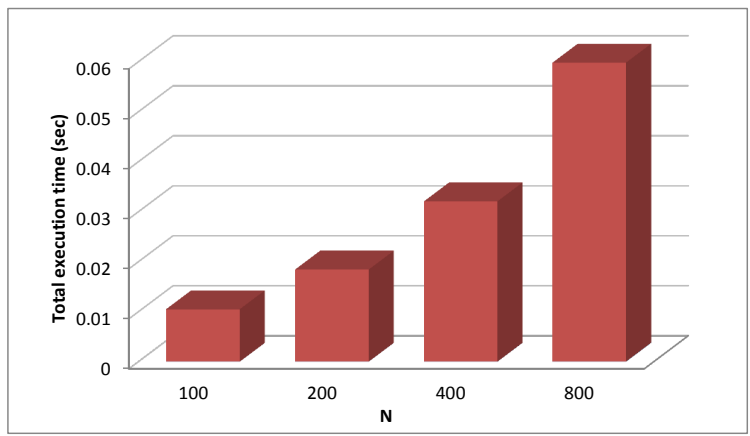}\vspace{-0.3in}
\caption{Total execution time for \textbf{\textit{asynchronous}} parallel computing algorithm with respect to variation in the number of grid points $N$}\label{fig.5}
\end{figure}

For comparative purposes, we carried out multiple simulations while increasing the number of grid points in 1D heat equation. Fig. \ref{fig.4} and \ref{fig.5} present the total execution time, which describes how much physical time has elapsed up to the given iteration steps. 
For both synchronous and asynchronous cases, the total execution time increases as the number of grid points $N$ increases. This is because the computation cost grows with proportional to the problem size. 
As evidently shown in Fig. \ref{fig.4} and \ref{fig.5}, the asynchronous algorithm drastically speeds up the execution time, and hence benefits the computing performance. In almost all cases, it is observed that asynchronous scheme outperforms synchronous scheme. Particularly when $N=100$, asynchronous scheme brought more than $30\times$ speedup compared to the synchronous scheme, which is substantial and outstanding acceleration.

\subsection{Intriguing Remarks}
The stability result for 1D heat equation with Dirichlet boundary condition is given in \citep{lee2015async}. In \citep{lee2015async}, we guaranteed that starting from given initial condition, temperature converges to unique steady-state distribution after sufficiently large iterations, \textit{regardless of} asynchrony. This implies that the steady-state solution obtained by asynchronous scheme is independent of asynchronous behavior even though it affects on the transient jitters. Note that the notion of stability in this case means that the temperature distribution does not diverge as well as converges to \textit{unique} steady-state value. 

\begin{figure}
\centering
\includegraphics[scale=0.6]{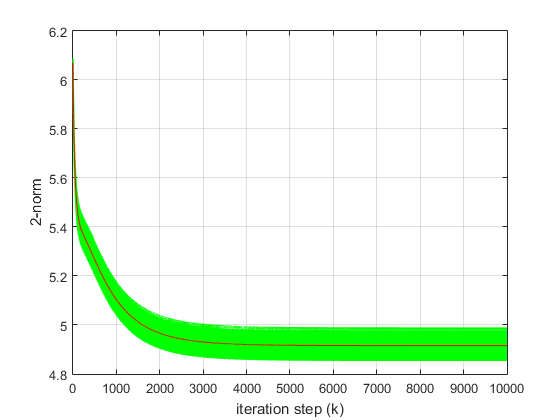}
\caption{The ensemble (green lines) of $300$ trajectories in 2-norm values and their corresponding mean value (red line) for \textit{\textbf{periodic}} 1D heat equation with implementation of asynchronous parallel computing algorithm.}\label{fig.6}
\end{figure}

However, 1D heat equation with \textit{periodic} boundary condition implemented in asynchronous scheme does not have unique steady-state solution. Here the periodic boundary condition denotes the case in which temperature at both end-points depends on each other, i.e.,
\begin{align*}
u_1(k) &= r(u_2(k) - 2u_1(k) + u_N(k)) + u_1(k),\\
u_N(k) &= r(u_1(k) - 2u_N(k) + u_{N-1}(k)) + u_N(k).
\end{align*}

The \textsf{CUDA} code below presents the asynchronous 1D heat equation with periodic boundary condition.

\begin{lstlisting}
__global__ void kernel(float* u, float* v){

    int i = blockIdx.x*blockDim.x + threadIdx.x;
  
    for(int k=0;k<kend;k++){
	if( i > 0 && i < N-1){
	   u[i] = r*(u[i+1]-2*u[i]+u[i-1]) + u[i];
	}
	/* periodic boundary condition */
	u[0] = r*(u[1]-2*u[0]+u[N-1]) + u[0];
	u[N-1] = r*(u[0]-2*u[N-1]+u[N-2]) + u[N-1];
	
	v[N*k+i] = u[i];
    }
}
\end{lstlisting}

In Fig. \ref{fig.6}, we demonstrate multiple simulations starting from same initial condition given by cosine function as in the previous case. The ensemble of 300 trajectories spreads out and does not converge to unique value, which is different from asynchronous 1D heat equation with Dirichlet boundary condition. Thus, in the case of periodic 1D heat equation, it is observed that asynchronous parallel computing algorithm is numerically stable (in the sense that the solution does not diverge), but the solution is not unique.
It is very interesting to see that even with exactly same PDE, same finite difference scheme, and same initial condition, one may expect different convergence results for different boundary conditions.
Note that the synchronous solution of periodic 1D heat equation always ends up with unique steady-state temperature distribution, since there is no randomness in synchronous scheme. This implies that incorrect information would be delivered 
by asynchronous scheme. In fact, the convergence or stability of asynchronous scheme is problem-dependent! (i.e., case by case). Therefore, asynchronous parallel computing algorithms give rise to a tradeoff issue between speedup and accuracy and hence, implementation of asynchronous parallel computing algorithms requires rigorous mathematical analysis before it is fully implemented.


\bibliography{mybibfile}

\end{document}